%% file: colm_conference.tex
\documentclass{article} 
\usepackage[preprint]{colm_conference}

\usepackage{amsmath}
\usepackage{microtype}
\usepackage{hyperref}
\usepackage{url}
\usepackage{booktabs}
\definecolor{darkblue}{rgb}{0, 0, 0.5}
\hypersetup{colorlinks=true, citecolor=darkblue, linkcolor=darkblue, urlcolor=darkblue}
\usepackage{tcolorbox}
\usepackage{colortbl}
\usepackage{multirow}
\usepackage{cleveref}
\usepackage{adjustbox}
\usepackage{arydshln}
\usepackage{pifont}
\usepackage{bbding}
\usepackage{fontawesome5}
\usepackage{pifont}
\usepackage{subcaption}
\usepackage{float}
\usepackage{tikz}
\usepackage{pgfplots}

\pgfplotsset{compat=1.18}

\newcommand{\ghidra}{Ghidra}
\newcommand{\cmark}{\ding{51}}
\newcommand{\xmark}{\ding{55}}
\newcommand{\crmark}{\textcolor{red}{\ding{51}}}

\newcommand{\toolemoji}{\faTools}
\newcommand{\labelemoji}{\faTags}

\title{ReF Decompile: Relabeling and Function Call Enhanced Decompile}

\author{Yunlong Feng \& BoHan Li  \\
Harbin Institute of Technology \\
\texttt{\{ylfeng,bhli\}@ir.hit.edu.cn}
\And
Xiaoming Shi \\
East China Normal University \\
\texttt{xmshi@cs.ecnu.edu.cn}
\And
Qingfu Zhu$^*$ \& Wanxiang Che\thanks{Corresponding author} \\
Harbin Institute of Technology \\
\texttt{\{qfzhu,car\}@ir.hit.edu.cn}
}

\begin{document}

\maketitle

\input{sections/abstract}
\input{sections/introduction.tex}
\input{sections/method.tex}

\input{sections/analysis.tex}
\input{sections/related.tex}
\input{sections/conclusion.tex}

\bibliography{main}
\bibliographystyle{colm_conference}

\end{document}

%% file: sections/abstract.tex

\begin{abstract}
The goal of decompilation is to convert compiled low-level code (e.g., assembly code) back into high-level programming languages, enabling analysis in scenarios where source code is unavailable.
This task supports various reverse engineering applications, such as vulnerability identification, malware analysis, and legacy software migration.  
The end-to-end decompile method based on large langauge models (LLMs) reduces reliance on additional tools and minimizes manual intervention due to its inherent properties.
However, previous end-to-end methods often lose critical information necessary for reconstructing control flow structures and variables when processing binary files, making it challenging to accurately recover the program's logic.
To address these issues, we propose the \textbf{ReF Decompile} method, which incorporates the following innovations:
(1) The Relabelling strategy replaces jump target addresses with labels, preserving control flow clarity.
(2) The Function Call strategy infers variable types and retrieves missing variable information from binary files.
Experimental results on the Humaneval-Decompile Benchmark demonstrate that ReF Decompile surpasses comparable baselines and achieves state-of-the-art (SOTA) performance of $61.43\%$.
The code and models has been released\footnote{\href{https://github.com/AlongWY/ReF-Dec}{https://github.com/AlongWY/ReF-Dec}}.
\end{abstract}

%% file: sections/introduction.tex
\section{INTRODUCTION}

Decompilation \citep{ghidra,idapro,llm4decompile,refine_decompile,feng2024self} is the reverse process of converting compiled binary code back into a high-level programming language, with the goal of recovering source code that is functionally equivalent to the original executable.
Decompilation has alluring application value in performing various reverse engineering tasks, such as vulnerability identification, malware analysis, and legacy software migration \citep{decompilation1,decompilation2_rnn,llm4decompile,feng2024self,nova}.

Despite the development of various rule-based decompilation tools, such as Ghidra \citep{ghidra} and IDA Pro \citep{idapro}, the decompilation process continues to face significant challenges. These include the inherent loss of information during compilation \citep{variable_name,for_loop} and the heavy reliance on manual effort to analyze and summarize assembly code patterns for rule-based approaches\citep{ghidra,idapro}. 
Moreover, uncovered or misinterpreted patterns can lead to inaccuracies in the decompilation results\citep{decompilation1,decompile_ir1,variable_name,for_loop,llm4decompile}.
To address this issue, researchers explore the use of large language models (LLMs) in decompilation tasks\citep{slade,btc,llm4decompile,nova,feng2024self,refine_decompile}.
By automatically learning the mapping patterns between low- and high-level code from aligned corpora, LLMs can reduce human labor and improve the readability of the generated output.

\begin{figure}[t]
    \centering
    \includegraphics[width=0.98\linewidth, alt={Illustration of neural network-based decompilation approaches}]{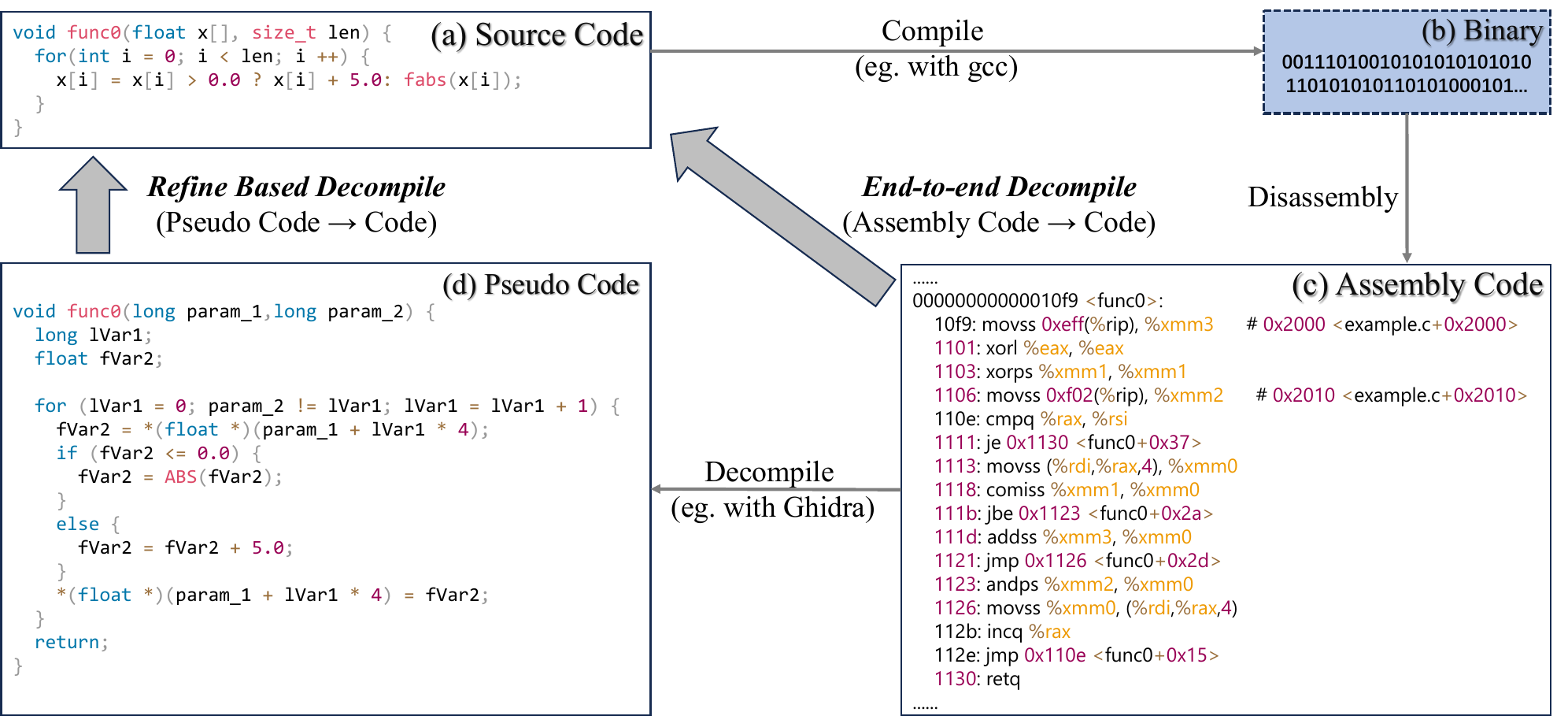}

    \caption{LLM-based decompilation methods can primarily be categorized into two types: (1) refine-based methods, which aim to refine the pseudo code generated by decompilers such as Ghidra to recover the original code; and (2) end-to-end methods, which aim to reconstruct the original code directly from assembly code.}
    \label{fig:intro}
\end{figure}

The LLM-based decompilation methods are typically divided into two categories: refine-based methods\citep{llm4decompile,refine_decompile,hu2024degpt} and end-to-end methods\citep{feng2024self,slade,btc,nova}.
As shown in \Cref{fig:intro}, the refine-based method focuses on recovering the original code (a) from the pseudo code (d) produced by existing decompilation tools, whereas the end-to-end method aims to directly reconstruct the original code (a) from assembly code (c).
Compared to the refine-based approach, the end-to-end method reduces reliance on additional tools and minimizes manual intervention due to its inherent properties.
This work focuses on improving end-to-end methods.

However, despite its advantages, previous end-to-end method encounters significant challenges in reconstructing control flow and variable values, limiting its accuracy and practical applicability. These limitations arise primarily from two issues:
(1) During data preprocessing, address information is directly and simplistically removed, making it difficult to recover control flow information from the processed assembly code; and 
(2) end-to-end methods rely solely on the processed assembly code, which lacks a significant amount of variable value information (e.g., floating-point numbers, strings, etc.).
These limitations hinder the accuracy and completeness of the decompiled results.

To address the above issues, we propose \textbf{Ref Decompile} method, which is designed to optimize the decompilation process of the end-to-end method.
(1) To tackle the problem of missing control flow information, the \textbf{\textit{Relabeling}} strategy is introduced to restructure the data format, which keeps the related information of jump instructions.
The processed result also satisfies the syntax accepted by the assembler.
(2) To solve the problem of missing variable information, the \textbf{\textit{Function call}} strategy provides a mechanism for the model to interact with the binary file to obtain variable values, thus completing the information needed to recover the variables.
By combining these two strategies, we leverage both control flow and variable information from the binary file, improving the precision of the decompiled results.

On the Humaneval-Decompile Benchmark, the Ref Decompile method outperforms the strongest baseline by $8.69\%$. It achieves a new state-of-the-art (SOTA) performance of $61.43\%$ and the highest readability score of $3.69$. 
These results demonstrate the effectiveness of the two strategies (Relabeling and Function Call) in improving the correctness and readability of decompiled outputs.

Our contributions are as follows:
\begin{itemize}
    \item We identify the key weaknesses of previous end-to-end methods: loss of control flow and variable information.
    \item To address these challenges, we redesign the end-to-end decompilation process, incorporating a \textit{relabeling} strategy to preserve control flow information and a \textit{function call} strategy to access variable information.
    \item The proposed Ref Decompile method achieves SOTA performance among models of the same size, surpassing the strongest baseline with $61.43\%$ in re-executability rate and achieving the highest readability score of $3.69$.
\end{itemize}

%% file: sections/method.tex
\section{METHODOLOGY}
\label{sec:method}

\begin{figure}[t]
    \centering
    \includegraphics[width=0.9\linewidth]{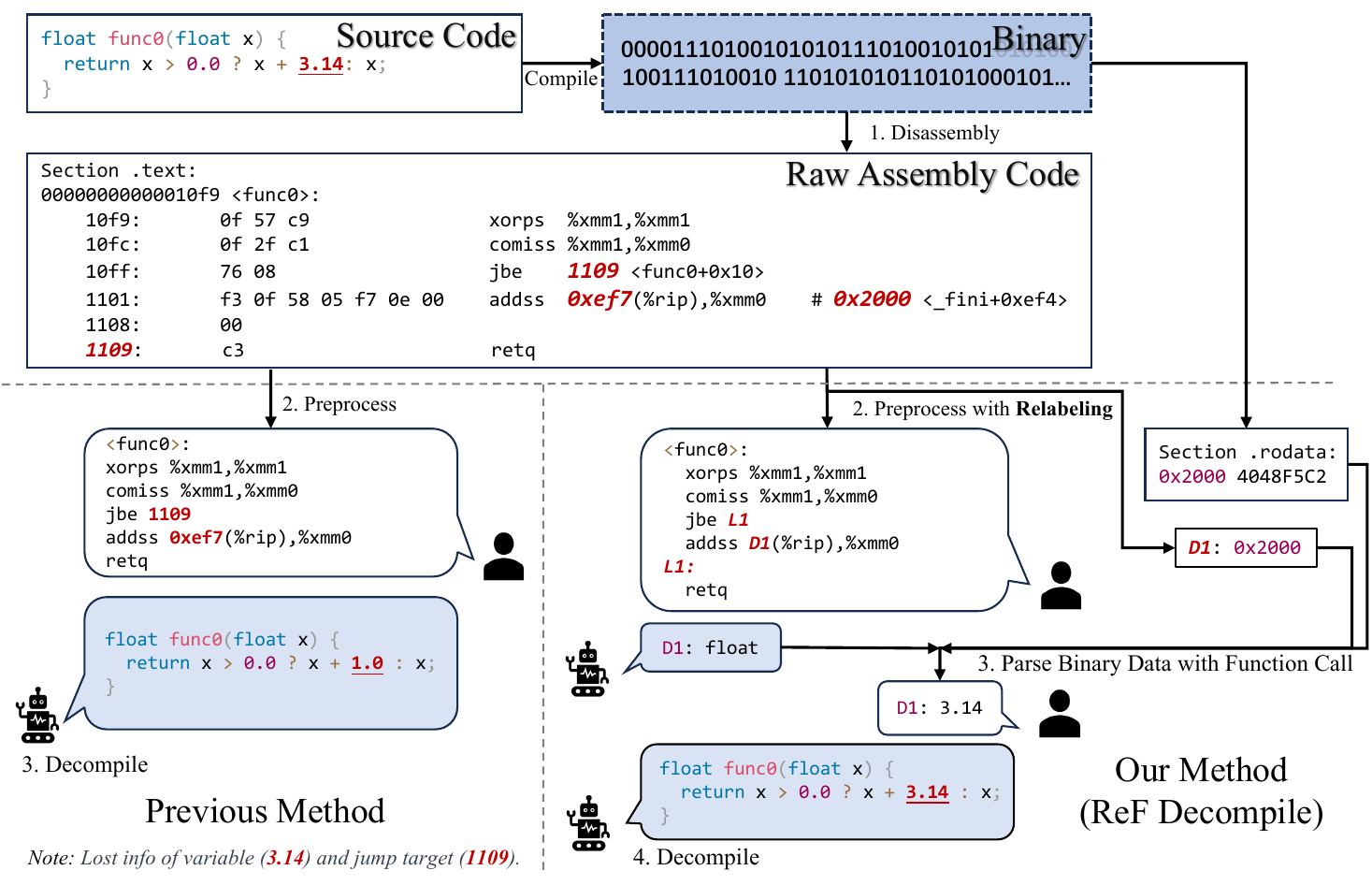}
    \caption{Comparsion of the previous method and our method (ReF Decompile). 
    Previous end-to-end methods rely solely on information from the executable segment, leading to ``information loss'' during decompilation. 
    For example, the processed assembly here lost variable information (“\textbf{\textit{3.14}}” in the source code) and the jump target (“\textbf{\textit{1109}}” of “jbe 1109” in the raw assembly).
    This results in code reconstructions that appeared plausible but are actually incorrect.
    By incorporating Relabeling information (Relabeling) and leveraging relevant tools (Function Call), the model can now gain a deeper understanding of code jump logic and access valuable information stored outside the executable segment. 
    This enhancement allows the model to accurately reconstruct the original code, significantly improving the precision and reliability of the decompilation process. }
    \label{fig:main-method}
\end{figure}

Existing end-to-end decompilation methods often lose critical information needed to reconstruct control flow structures and variables when processing binary files, making it difficult to accurately recover the program’s logic.  
For example, in \Cref{fig:main-method}, the processed assembly of the previous method lost variable information (``$3.14$'' in the source code) and the jump target (``1109'' of ``jbe 1109'' in the raw assembly).

To solve the above challenge, in this paper, we propose the Relabeling and Function Call Enhanced Decompile method (ReF Decompile), which updates and optimizes the end-to-end decompilation process and applies it to both the training and inference stages.
The method includes using the Relabeling strategy to identify address information, and leveraging the function call strategy to complete variable information using the binary file.

In this section, we first introduce our proposed ReF Decompile and then describe how the training data are constructed.

\subsection{ReF Decompile: Relabeling and Function Call Enhanced Decompile}

In this subsection, we will introduce the overall process of the ReF Decompile method, which consists of the following four steps:

\begin{enumerate}
    \item \text{Disassembly Binary File:}
    By employing disassembly tools such as ``Objdump'' or ``Capstone'', and by specifying concrete function names or address ranges, we can translate the machine code of functions into raw assembly code that is more interpretable.
    Raw assembly code includes the complete source code information that will be used in the following steps.

    \item \text{Preprocess Assembly with Relabeling:}
    In this phase, we preprocess the raw assembly code to simplify its structure for data construction. 
    The processed result also satisfies the syntax accepted by the assembler.
    Initially, all jump instruction target addresses are replaced with labels (e.g., address 0x1109 is replaced with "L1"), and these labels are inserted at the corresponding target locations while removing all address information preceding non-target addresses. 
    Secondly, addresses in instructions related to data access are also labeled (e.g., \textit{0xef7(\%rip)} is replaced with \textit{D1(\%rip)}), and a mapping between these labels and their actual addresses is recorded.
    The preprocessed assembly code is then utilized as input for the model.

    \item \text{Process Function Call Request:}
    Our system incorporates a mechanism that allows the model to generate structured requests for function calls or accessing data.
    Upon feeding the preprocessed assembly code into the model, if it contains memory access instructions, the model typically issues a function call request that includes labels from the memory access instructions along with their associated data types.
    Leveraging the label-to-address mapping established in the previous step, we can identify the actual addresses represented by these labels and parse the corresponding sections of the binary file (commonly the .rodata section) according to the data types predicted by the model. The labels and parsing results are then fed back into the model as input.

    \item \text{Decompile Finally:}
    After completing the aforementioned steps, the model accumulates both the assembly code corresponding to the function and the data values at the addresses referenced by memory access instructions. Based on this information, the model produces a final high-level language representation, thereby achieving the transformation from low-level machine code to high-level programming language constructs.

\end{enumerate}

\subsection{Data construction}

In this section, we introduce the data construction process of \textit{Relabeling} and \textit{Function Call} in our proposed method.
The constructed data trains the model to understand the format of relabeled data and to interact through function calls.

\begin{figure}[t]
    \centering
    \includegraphics[width=0.9\linewidth]{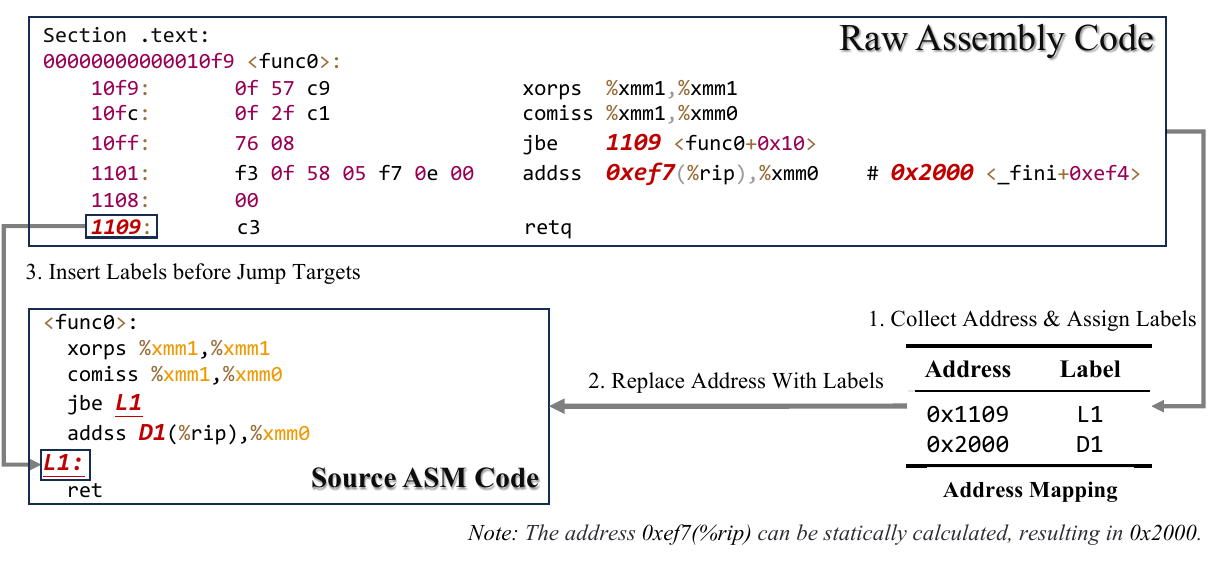}
    \caption{The processing details of Relabeling.}
    \label{fig:relabeling}
\end{figure}

\subsubsection{Relabeling}

The relabeling process aims to remove specific address information (including jump and memory access addresses) from the assembly code while preserving the program's jump logic to ensure control flow integrity. 
As the \Cref{fig:relabeling} shows, the specific steps are as follows:

\begin{enumerate}
    \item \textbf{Collection of Address Information and Label Assignment}: First, we perform a disassembly analysis of the program. (a) We identify all target addresses associated with jump instructions (e.g., ``jbe''), as these addresses control the program's flow of execution and are crucial for the relabeling process. These addresses are recorded and labeled (e.g., L1, L2, etc.). (b) We identify all instructions related to memory access (e.g., ``addss 0xef7(\%rip),\%xmm0''), record these addresses, and assign corresponding labels (e.g., D1, D2, etc.).
    \item \textbf{Replacement of Specific Addresses with Labels}: All addresses recorded in the previous step, including both jump and memory access addresses, are replaced with the assigned labels.
    \item \textbf{Insertion of Labels Before Jump Targets}: To eliminate specific memory address information while maintaining the jump logic, we insert the corresponding label before each jump target instruction.
\end{enumerate}

Through this process, we remove the address information while preserving the integrity of the program's jump logic and ensuring that critical jump information is not lost.

\subsubsection{Function Call}

This process is designed to lay the groundwork for subsequent decompilation efforts using LLMs.
This involves parsing source code and analyzing binary files to accurately extract and match literals with their storage addresses within the binaries. 
As the \Cref{fig:tool} shows, the detailed process is as follows:

\begin{figure}[t]
    \centering
    \includegraphics[width=0.9\linewidth]{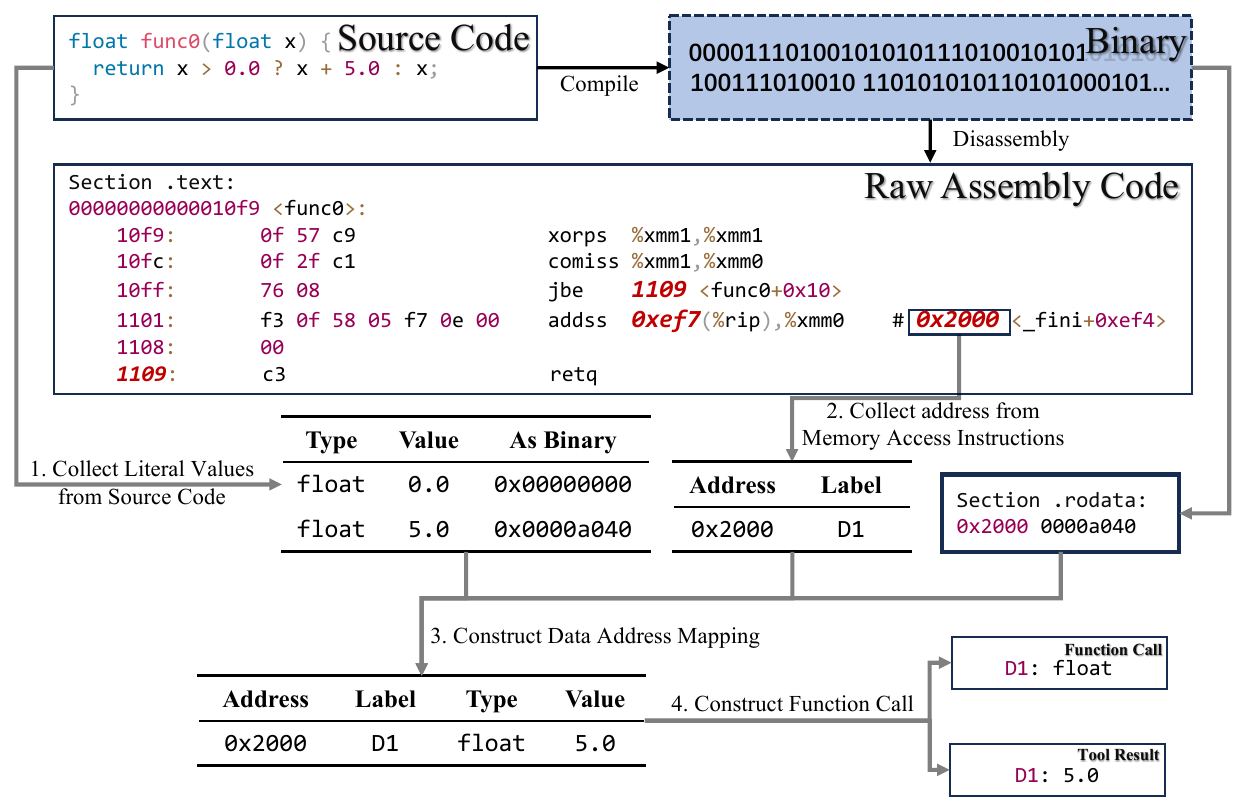}
    \caption{Overview of the Data Construction for Function Call. }
    \label{fig:tool}
\end{figure}

\begin{enumerate}
\item \textbf{Collect Literals from Source Code:}
Initially, we employ the clang compiler toolchain to parse C language code, extracting literal data such as strings, floats, and doubles. This step identifies all constants within the source code, clarifying how these constants manifest post-compilation. To ensure consistency between literals in the source code (for example, the values 0.0 and 5.0 in the source code in Figure 4) and data in the binary files, we convert the extracted literals into corresponding bytecode based on their types. This conversion ensures that the storage format of literals matches the form present in the binary files, facilitating accurate comparison.

\item \textbf{Collecting Memory Access Addresses:}
Next, we disassemble the compiled binary files to identify memory addresses related to memory access instructions, assigning labels (e.g., D1, D2) to these addresses. This step mirrors the relabeling process, specifically focusing on address mapping pertinent to memory accesses.

\item \textbf{Constructing Data Addresses Mapping:}
Finally, we compare the bytecode representation of literals with the memory load addresses obtained from the disassembly analysis. Through this precise matching, we determine the specific storage addresses and types of literals within the binary files, thereby constructing a mapping relationship among labels, addresses, types, and data.

\item \textbf{Constructing Function Calls:}
Based on the established mapping relationships, we can formulate the necessary information for function calls, including the memory access labels and data types required by the model. The result of such a call is the value stored at the address corresponding to the label.
\end{enumerate}

This systematic approach provides high-quality input data for subsequent decompilation tasks using LLMs. By accurately extracting and matching literals with their storage addresses, and constructing function calls based on this mapping, the model gains a better understanding of program execution details, leading to more precise code representation generation.

%% file: sections/analysis.tex

\section{EXPERIMENTS}

In this section, we introduce the experimental details and analysis of our results. 
Following \citet{llm4decompile} and \citet{feng2024self}, we utilize a subset of Exebench \citep{exebench} as the training set and Decompile-Eval \citep{llm4decompile} as the test set.
We compare our method with six baseline methods and experimental results demonstrate that our method achieves state-of-the-art performance among models of the same size.

\subsection{Training Details}



\subsubsection{Training Data}
Following \citet{llm4decompile} and \citet{feng2024self}, we utilize a subset of Exebench \citep{exebench} as the training set.
ExeBench is the largest public collection of five million C functions, and we select 15k samples from the train\_real\_compilable subset to synthesis the training data (about 0.4b tokens). The selected functions exclusively utilize the standard C library and do not include additional data structures. The training data were synthesized with gcc 11.4 provided by Ubuntu 22.04.

\subsubsection{Implementation} 

Following \citet{feng2024self}, we employ LoRA \citep{hu2022lora} to fine-tune the \textit{\href{https://huggingface.co/LLM4Binary/llm4decompile-6.7b-v1.5}{llm4decompile-6.7b-end v1.5}} model obtained from Hugging Face \citep{huggingface}.
The rank is set to 32, alpha to 64, and the target includes embedding layer, lm head, and all projection layers\footnote{embed\_tokens, lm\_head, q\_proj, k\_proj, v\_proj, o\_proj, gate\_proj, up\_proj, down\_proj}.
The model is trained for one epoch using the AdamW optimizer \citep{adamw} with a learning rate of 5e-5.
The maximum sequence length is set to 4096, and the learning rate scheduler type is cosine, with a warm-up period of 20 steps. 
The fine-tuning process leverages LlamaFactory \citep{llamafactory}, FlashAttention 2 \citep{flashattention2}, and DeepSpeed \citep{wang2024zero}. All experiments are conducted on an A100-SXM4-80GB GPU, and greedy decoding is utilized throughout the experiments.


\subsection{Evaluation Details}


\subsubsection{Benchmark}

Following \citet{llm4decompile} and \citet{feng2024self}, we employ Decompile-Eval \citep{llm4decompile} as our evaluation benchmark, which is specifically designed to assess the decompilation capabilities of large language models.
The Decompile-Eval benchmark\citep{llm4decompile} is adapted from the HumanEval benchmark\citep{humaneval}, which includes 164 problems initially designed for code generation tasks. These problems are translated into the C programming language, and the corresponding assembly code is generated at four optimization levels (O0, O1, O2, and O3). The correctness of the decompilation results is tested using the test cases from HumanEval.

\subsubsection{Metrics}
The primary metrics of the Decompile-Eval benchmark are as follows:

\begin{itemize}
  \item \textbf{Re-executability Rate}: This metric assesses the functional correctness of the decompiled code. Specifically, it measures whether the recompiled binaries produce the expected outputs when executed. The correctness of the output is determined using the testing methodology provided by the HumanEval dataset, ensuring a comprehensive evaluation of the logical accuracy of the decompiled code.
  \item \textbf{Readability}: This metric evaluates the readability of the decompiled code. Specifically, it uses GPT-4o with a structured template to assess syntactic similarity (variables, loops, conditions) and structural integrity (logical flow, overall structure). Based on a detailed comparison between the original and decompiled code, a score from 1 (Poor) to 5 (Excellent) is assigned. A score of 4 indicates that the decompiled code is nearly identical to the original in terms of readability, offering an intuitive measure of code quality.
\end{itemize}



\subsection{Baselines}
\label{sec:baselines}

To demonstrate the effectiveness of our proposed methods, we compare them with several baselines, including Rule-Based Decompilers, Refine-Based Methods, and End-to-End Methods.
In this section, we introduce these methods.\footnote{For the baselines listed below, all models are assumed to be of size 6.7B unless otherwise specified, except for \ghidra{} and GPT-4o. We also report the performance of models with other sizes in \Cref{fig:compare}}


\begin{itemize}
    \item \textbf{Rule-Based Decompiler} relies on manually crafted rules and techniques such as control flow and data flow analysis to transform assembly code into high-level language code. 
    \begin{itemize}
    \item \textit{\ghidra{}}: A free and open-source reverse engineering tool (decompiler)
    \citep{ghidra}. It serves not only as the baseline for comparison but also as the preprocessing tool for the Refine-Based decompilation method.
    \end{itemize}
    \item \textbf{Refine-Based Methods} builds upon the output of rule-based decompilers, leveraging large models to refine and enhance the decompilation results for improved accuracy and readability.
    \begin{itemize}
    \item \textit{GPT 4o}: One of the most powerful language models developed by OpenAI, which is used to refine the Ghidra decompilation output.
    \item \textit{LLM4Decompile-Ref}: A series of pre-trained refine-based models from LLM4Decompile~\citep{llm4decompile}, which refine pseudo-code decompiled by Ghidra.
    \end{itemize}
    \item \textbf{End-to-End Methods} directly process assembly code using large models to generate high-level language code.
    \begin{itemize} 
    \item \textit{LLM4Decompile-End}: A series of pre-trained end-to-end models from LLM4Decompile~\citep{llm4decompile}, which directly decompile binaries into high-level code.
    \item \textit{FAE Decompile}: A model obtained by applying the Fine-grained Alignment Enhancement method to further fine-tune the llm4decompile-End-6.7b \citep{feng2024self}\footnote{This paper also involves a decompilation strategy called SC$^2$. We do not include it as a baseline since it is not a model.}.
    
    \end{itemize}
\end{itemize}

\subsection{Main Results}


In this paper, we evaluate our ReF Decompile and the backbones in \Cref{sec:baselines} on Decompile-Eval. The main results of our experiments are shown in ~\Cref{table:main} and ~\Cref{fig:compare}. 

\input{tables/main.tex}
\input{figures/compare}


\textbf{(1) Our method is the best and surpasses all other approaches to become the state of the art.} As shown in \Cref{table:main}, the decompiler-based method has the lowest performance, with a re-compilability of 20.12, because manually crafted rule systems cannot guarantee that the generated code is fully compilable. In the refine-based methods, LLM4decompile, which is trained on 20B tokens of decompilation data, outperforms the untrained GPT-4o, achieving a re-compilability of 52.74\%. As analyzed in the introduction, it corrects the decompilation results from the decompiler, surpassing the two other end-to-end methods, and becomes the strongest baseline.

\textbf{(2) As an end-to-end approach, ReF Decompile surpasses refine-based baselines.} It improves the Re-executability metric by 8.69\% and the readability metric by 0.19. This demonstrates the effectiveness of our two strategies, Relabling and Function Call, which reverse the trend where end-to-end methods typically perform worse than refine-based methods.

\textbf{(3) The Relabling and Function Call strategies better leverage the potential of end-to-end methods.} As shown in \Cref{fig:compare}, the performance of 6.7B ReF Decompile not only significantly surpasses both 1.3B and 6.7B Refine-Based LLM4Decompile-Ref, but it is also comparable to the 22B Refine-Based LLM4Decompile-Ref, with an average gap of only 2.75\%. Notably, at optimization level O0, the performance of ReF Decompile (6.7B) even exceeds that of the 22B Refine-Based LLM4Decompile-Ref model, indicating that the model can automatically learn patterns beyond those defined by humans from large-scale corpora.

\textbf{(4) ReF Decompile surpasses other end-to-end baselines in readability, becoming the new SOTA.} Besides a significant improvement in Re-executability (10.36\%), it achieves a readability score of 3.69. This shows that our two strategies effectively avoid the loss of crucial information needed to reconstruct control flow structures and variables, leading to a more accurate recovery of the program's logic.

%


\subsection{Ablation Study}

\input{tables/ablation.tex}

As shown in \Cref{tab:ablation}, we conduct ablation experiments to analyze the impact of two different strategies (Relabeling and Function Call) on model performance.

When initialized with the LLM4Decompile-End-6.7B model, both Relabeling and Function Call contribute to model performance and readability, and their combination performs best in both metrics.
Specifically, the introduction of Relabeling improves performance by 3\%, while the introduction of Function Call improves performance by 7\%. The model with both Relabeling and Function Call achieves the highest average re-executability at 61.43\%. Moreover, these two components also improve the readability of the decompiled results, increasing it by about 0.14 points, which corresponds to a 4\% improvement in readability.

In addition to using LLM4Decompile-End-6.7B, in order to minimize the impact of continued pretraining, we also use its initialized model Deepseek-Coder-6.7B-base. Surprisingly, the model with Relabeling and Function Call, trained with only 0.4B tokens, outperforms the strongest baseline of the same size in \Cref{table:main} (52.74). The ablation results further confirm that Relabeling and Function Call both lead to significant performance improvements, with the combined effect being even more pronounced. Specifically, the introduction of Relabeling leads to a 3\% performance improvement, Function Call leads to a 5\% improvement, and the simultaneous introduction of both components leads to a 10\% improvement, surpassing the combined performance improvements of each component individually. Similar to the model initialized with LLM4Decompile-End-6.7B, both components also improve the readability of the decompiled results by about 0.14.



\subsection{Analysis of Two Components for Untuned Models}

This section analyzes the impact of two key components—Relabeling and Tool Integration—on the performance of untuned models. We examine how these strategies enhance decompilation accuracy.

\input{tables/ablation-untune.tex}



\subsubsection{Relabeling improves the performance of untuned models significantly:}
As shown in \Cref{tab:untuned-ablation}, Relabeling enhances the readability of jump addresses in assembly code by assigning more intuitive labels, leading to better decompilation results.
For example, performance on the GPT-4o~\citep{gpt4} model improves from 21.34\% to 27.74\% (a gain of 6.40\%), while on the Qwen model~\citep{qwen2.5coder}, it improves from 11.28\% to 14.63\% (a gain of 3.37\%).
This supports our hypothesis that Relabeling makes assembly code’s jump logic easier for models to understand, resulting in better inference. GPT-4o benefits more from this change, likely due to its stronger ability to handle complex logic.
Relabeling demonstrates how simple preprocessing improvements can significantly boost performance without requiring additional fine-tuning. 
Enhancing input readability and logic clarity proves to be a valuable strategy for improving model effectiveness in specific tasks.


\subsubsection{Potential of Function Call to Enhance Untuned Model Performance Remains Underexplored:}
As shown in \Cref{tab:untuned-ablation}, introducing tools does not always lead to performance improvements.
In our experiments, although we provide the model with tools for decompilation in the prompt, it consistently returns the decompiled results directly without invoking any tools.
For GPT-4o, Function Call slightly reduces performance. 
This decrease might be due to the model's inability to determine when to invoke tools to retrieve data from unknown addresses.
Additionally, prompts related to tool usage may cause interference.
In contrast, in the Qwen experiments, Function Call slightly improves performance.
Although the model does not actively do function call, the descriptive prompts associated with tools may stimulate some decompilation capabilities. 

To further verify the model's ability to utilize function call information, we simulate function call and responses to test whether the model can use tool outputs. 
The results show that models, despite not calling tools themselves, are able to leverage the provided information to improve performance. 
For example, tools supply variable information missing from the assembly code, allowing models to avoid "guessing" variable values. 
This leads to a performance boost of 2\% to 5\% for GPT-4o.

These findings highlight two key points. 
First, current models struggle to invoke and utilize tools effectively, limiting their immediate benefits. 
Second, tools still show significant potential for performance enhancement, as demonstrated by the gains from simulated tool use. 

\input{tables/dataset-rate.tex}
\input{tables/dataset.tex}

\subsection{Dataset Analysis}
In this section, to validate the necessity of Relabelling and Function Call, we analyze the distribution of information related to these two strategies within the dataset.


\subsubsection{Rodata is widely present in various environments:} 
To demonstrate the necessity of introducing function calls, we analyze the Rodata information in Exebench and Humaneval-Decompile in \Cref{tab:dataset-rate}.
In Humaneval-Decompile, 27.90\% of the code includes data within the rodata segment.
Similarly, in Exebench's train\_real\_compilable subset, 48\% of the code stores certain information in the rodata section rather than the executable code section after compilation.
It is commonly believed that Exebench approximates the distribution of real-world code scenarios, while Humaneval is relatively simple. 
Thus, we can observe that even in relatively simple scenarios like Humaneval, nearly 1/4 of the code includes some rodata segment data. 
This implies that in real-world scenarios, the introduction of function call enables models to read the contents of the rodata segment, potentially leading to greater benefits.

\subsubsection{Relabelling and Function Call Are Widely Applicable:}

As shown in \Cref{tab:dataset-rate}, more than 80\% of the code includes fixed addresses, such as data memory access addresses and jump targets, demonstrating the broad applicability of the Relabelling strategy. 
Decompile-Eval contains more branch and jump instructions (Jump Instructions), whereas Exebench exhibits more frequent access to rodata (Data Labels and Load Instructions).

As shown in \Cref{tab:dataset}, at the O2 optimization level, both datasets exhibit the lowest numbers of memory access and jump instructions, which could explain the relatively smaller performance gains from Relabelling and Function Call at this level. 
Nonetheless, our methods demonstrate significant improvements across other optimization levels, underscoring their robustness and adaptability in diverse optimization settings.



%% file: tables/main.tex
\setlength{\dashlinedash}{0.5pt}
\setlength{\dashlinegap}{1.5pt}
\setlength{\arrayrulewidth}{0.3pt}

\begin{table*}
    [t]
    \centering
    \begin{adjustbox}{width=\textwidth}
        \begin{tabular}{lcccccccccc}
            \toprule \multirow{2}{*}{\textbf{Model/Metrics}}                                              & \multicolumn{5}{c}{Re-executability Rate (\%)} & \multicolumn{5}{c}{Readability (\#)} \\
            \cmidrule(lr){2-6} \cmidrule(lr){7-11}                                                        & O0                                        & O1                                 & O2             & O3             & AVG            & O0             & O1             & O2             & O3             & AVG            \\
            \hline
            \rowcolor[rgb]{0.93,0.93,0.93}\multicolumn{11}{c}{\textbf{Rule Based Decompiler}} \\
            \ghidra{}                                                                                     & 34.76                                     & 16.46                              & 15.24          & 14.02          & 20.12          & 2.98 & 2.41 & 2.52 & 2.38 & 2.57           \\
            \rowcolor[rgb]{0.93,0.93,0.93}\multicolumn{11}{c}{\textbf{Refine-Based Method}} \\
            GPT-4o                                                                           & 46.95                                     & 34.15                              & 28.66          & 31.10          & 35.22          & 2.82 & 2.35 & 2.29 & 2.31 & 2.44           \\
            LLM4Decompile-Ref                                                           & \underline{74.39}                                     & 46.95                              & \underline{47.56}          & \underline{42.07}          & \underline{52.74}          & \underline{4.08} & 3.38 & 3.34 & 3.19 & 3.50 \\
            \rowcolor[rgb]{0.93,0.93,0.93}\multicolumn{11}{c}{\textbf{End-to-End Method}} \\
            LLM4Decompile-End                                                                        & 69.51                                     & 44.51                              & 39.63          & 38.41          & 48.02          & 4.07 & \underline{3.46} & \underline{3.40} & 3.23 & \underline{3.54} \\
            FAE Decompile                                                                            & 67.68                                     & \underline{48.78}                              & 45.73          & 42.07          & 51.07          & 3.94 & \underline{3.46} & \underline{3.40} & \underline{3.25} & 3.51 \\
            \textbf{ReF Decompile}                                                                   & \textbf{85.37}                            & \textbf{56.10}                     & \textbf{51.83} & \textbf{52.43}          & \textbf{61.43} & \textbf{4.13} & \textbf{3.60} & \textbf{3.54} & \textbf{3.49} & \textbf{3.69} \\
            \bottomrule
        \end{tabular}
    \end{adjustbox}
    \caption{Main comparison of different approaches and models for re-executability rate
    and readability across different optimization levels (O0, O1, O2, and O3) on HumanEval-Decompile benchmark. \textbf{Bold} denotes the best performance.
    \underline{Underline} denotes the second-best performance.
    }
    \label{table:main}
\end{table*}

%% file: figures/compare.tex
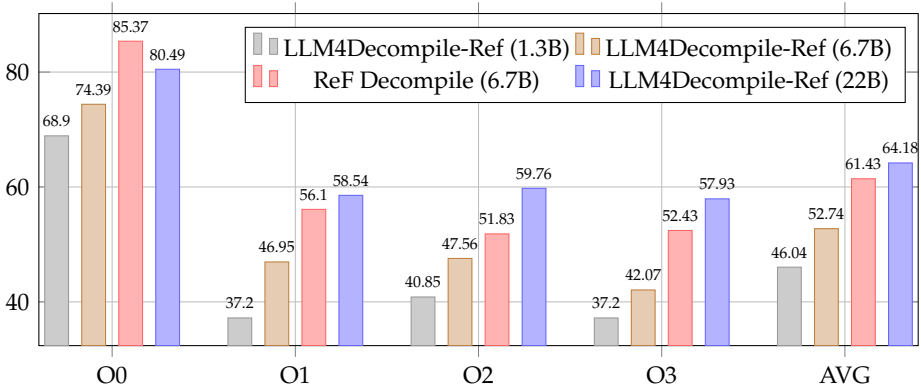
\begin{figure}[t]
 \centering
 \begin{tikzpicture} 
 \begin{axis}[
 enlargelimits=0.10,
 legend style={at={(0.61,0.96)},
  anchor=north,legend columns=2},
 symbolic x coords={O0, O1, O2, O3, AVG},
 xtick=data,
 ybar=5pt,
 bar width=9pt,
 width=0.95\linewidth, height=6cm,
 nodes near coords,
 nodes near coords align={vertical},
 nodes near coords style={font=\tiny},
 font=\small,
 grid=major,
 ]
 \addplot[fill=black!20!white,draw=black!40!white] coordinates {
  (O0, 68.9)
  (O1, 37.2)
  (O2, 40.85)
  (O3, 37.2)
  (AVG, 46.04)
 };
 \addplot [fill=brown!40!white,draw=brown] coordinates {
  (O0, 74.39)
  (O1, 46.95)
  (O2, 47.56)
  (O3, 42.07)
  (AVG, 52.74)
 };
  \addplot [fill=red!30!white,draw=red!60!white] coordinates {
  (O0, 85.37)
  (O1, 56.1)
  (O2, 51.83)
  (O3, 52.43)
  (AVG, 61.43)
 };
  \addplot [fill=blue!30!white,draw=blue!60!white] coordinates {
  (O0, 80.49)
  (O1, 58.54)
  (O2, 59.76)
  (O3, 57.93)
  (AVG, 64.18)
 };
 \legend{LLM4Decompile-Ref (1.3B), LLM4Decompile-Ref (6.7B), ReF Decompile (6.7B), LLM4Decompile-Ref (22B)}
 \end{axis}
 \end{tikzpicture}
 \caption{Re-executability rate comparison between ReF Decompile and LLM4Decompile-Ref models of varying sizes (1.3B, 6.7B, and 22B parameters).}\label{fig:compare}
\end{figure}

%% file: tables/ablation.tex
\begin{table*}
  [t]
  \centering
  \resizebox{\textwidth}{!}{%
  \begin{tabular}{cccccccccccc}
    \toprule \multicolumn{2}{c}{Components}                                                             & \multicolumn{5}{c}{Re-executability Rate (\%)} & \multicolumn{5}{c}{Readability (\#)} \\
    \cmidrule(lr){1-2} \cmidrule(lr){3-7} \cmidrule(lr){8-12}        \hspace{8pt}\labelemoji\hspace{8pt}                                                                & \hspace{8pt}\toolemoji\hspace{8pt}                                      & O0                                 & O1             & O2             & O3             & AVG            & O0             & O1             & O2             & O3             & AVG            \\
    \hline
    \rowcolor[rgb]{0.93,0.93,0.93}\multicolumn{12}{c}{\textbf{Initialize with LLM4Decompile-End-6.7B~\citep{llm4decompile}}}   \\
    \xmark                                                                                              & \xmark                                    & 69.51                              & 46.95          & 50.61          & 46.34          & 53.35          & 3.98 & 3.41 & 3.44 & 3.38 & 3.55 \\
    \cmark                                                                                              & \xmark                                    & 75.61                              & 50.61          & 50.00          & 50.00          & 56.55          & 4.01 & 3.44 & 3.39 & \textbf{3.49} & 3.58 \\
    \xmark                                                                                              & \cmark                                    & 83.54                     & \textbf{56.10}          & 51.22          & 50.61 & 60.37 & 4.05 & 3.51 & 3.51 & 3.42 & 3.62 \\
    \cmark                                                                                              & \cmark                                    & \textbf{85.37}                            & \textbf{56.10}                     & \textbf{51.83} & \textbf{52.43}          & \textbf{61.43} & \textbf{4.13} & \textbf{3.60} & \textbf{3.54} & \textbf{3.49} & \textbf{3.69} \\

    \rowcolor[rgb]{0.93,0.93,0.93}\multicolumn{12}{c}{\textbf{Initialize with Deepseek-Coder-6.7B-base~\citep{deepseekcoder}}} \\
    \xmark                                                                                              & \xmark                                    & 59.15                              & 35.98          & 39.02          & 37.80          & 42.99          & 3.71 & 3.05 & 3.16 & 3.05 & 3.24 \\
    \cmark                                                                                              & \xmark                                    & 66.46                              & 41.46          & 38.41          & 36.59          & 45.73          & 3.76 & 3.17 & \textbf{3.21} & 3.08 & 3.31 \\
    \xmark                                                                                              & \cmark                                    & 70.73                              & 39.63          & 39.02          & 40.24          & 47.41          & 3.90 & 3.17 & 3.08 & 3.11 & 3.31 \\
    \cmark                                                                                              & \cmark                                    & \textbf{79.88}                     & \textbf{45.73} & \textbf{43.90} & \textbf{42.68} & \textbf{53.05} & \textbf{3.96} & \textbf{3.21} & 3.18 & \textbf{3.19} & \textbf{3.38} \\
    \bottomrule
  \end{tabular}%
  }
  \caption{The ablation study of different methods across four optimization levels
  (O0, O1, O2, O3), as well as their average scores (AVG). The results in bold represent the optimal performance. The ~\labelemoji~ and ~\toolemoji~ means Relabedling and Function Call. \textbf{Bold} denotes the best performance.}
  \label{tab:ablation}
\end{table*}

%% file: tables/ablation-untune.tex
\begin{table*}
  [t]
  \centering
  \resizebox{\textwidth}{!}{%
  \begin{tabular}{cccccccccccc}
    \toprule \multicolumn{2}{c}{Components}                                                             & \multicolumn{5}{c}{GPT-4o} & \multicolumn{5}{c}{Qwen2.5-Coder-32B-Instruct} \\
    \cmidrule(lr){1-2} \cmidrule(lr){3-7} \cmidrule(lr){8-12} \hspace{6pt}\labelemoji\hspace{6pt}                                                                & \hspace{6pt}\toolemoji\hspace{6pt}                                      & O0                                 & O1             & O2             & O3             & AVG            & O0             & O1             & O2             & O3             & AVG            \\
    \midrule
\xmark                           & \xmark  & 30.49          & 17.68          & 18.90          & 18.29          & 21.34          & 12.20          & 12.20          & 11.59          & 9.15           & 11.28          \\
\cmark                           & \xmark  & 37.80 & 26.22 & 24.39          & 22.56          & 27.74          & 15.24          & 16.46          & 14.02          & 12.80          & 14.63          \\
\xmark                           & \cmark  & 28.66          & 15.85          & 14.63          & 17.07          & 19.05          & 23.17          & 14.02          & 15.24          & 10.37          & 15.70          \\
\cmark                           & \cmark  & 31.71          & 25.61          & 22.56          & 25.00          & 26.22          & 23.78          & 16.46          & \textbf{17.07} & 10.37          & 16.92          \\ \hdashline 
\xmark & \crmark & 35.98 & 21.95          & 16.46          & 18.29          & 23.17          & \textbf{25.00} & 15.85          & 16.46          & \textbf{10.98} & \textbf{17.07} \\
\cmark & \crmark & \textbf{43.29} & \textbf{29.27} & \textbf{26.83} & \textbf{28.66} & \textbf{32.01} & 24.39 & \textbf{17.07} & \textbf{16.46} & 10.37 & \textbf{17.07} \\ \bottomrule
  \end{tabular}%
  }
  \caption{The analysis study of different components across four optimization levels (O0, O1, O2, O3), as well as their average scores (AVG). This study aims to investigate the performance impact of various components on the model without any fine-tuning. The symbol \crmark~ in the "Tool" column indicates that we have constructed dialogue turns involving tool usage, thereby compelling the model to refer to the outcomes of tool invocations. The ~\labelemoji~ and ~\toolemoji~ means Relabedling and Function Call. \textbf{Bold} denotes the best performance.}
  \label{tab:untuned-ablation}
\end{table*}

%% file: tables/dataset-rate.tex
\begin{table}[t]
\centering
\begin{tabular}{lrrrrrrrr}
\toprule
\multirow{2}{*}{Proportion Rate (\%)} & \multicolumn{4}{c}{Decompile-Eval}  & \multicolumn{4}{c}{Exebench} \\ \cmidrule(lr){2-5} \cmidrule(lr){6-9}
                         & O0     & O1     & O2     & O3     & O0       & O1      & O2      & O3     \\ \midrule
With Data Labels         & 25.61  & 26.21  & 25.00  & 34.76  & 47.61    & 47.62   & 47.75   & 49.01   \\
With Jump Labels         & 96.34  & 95.73  & 95.73  & 95.73  & 76.66    & 72.19   & 71.57   & 71.20   \\ 
With Both                & 25.00  & 25.61  & 24.39  & 34.15  & 37.67    & 36.74   & 36.81   & 37.88   \\
With Any                 & 96.95  & 96.34  & 96.34  & 96.34  & 86.61    & 83.06   & 82.51   & 82.33   \\  \bottomrule
\end{tabular}
  \caption{This table lists shows the proportion of samples with different labels under different optimization levels.}
  \label{tab:dataset-rate}
\end{table}

%% file: tables/dataset.tex
\begin{table}[t]
\centering
\begin{tabular}{lrrrrrrrr}
\toprule
\multirow{2}{*}{Mean Num (\#)} & \multicolumn{4}{c}{Exebench}  & \multicolumn{4}{c}{Decompile-Eval} \\ \cmidrule(lr){2-5} \cmidrule(lr){6-9}
                         & O0    & O1    & O2    & O3    & O0      & O1     & O2     & O3     \\ \midrule
Data Labels              & 2.46  & 2.44  & 2.43  & 2.49  & 0.80    & 0.85   & 0.78   & 1.10   \\
Jump Labels              & 4.13  & 3.62  & 3.49  & 3.81  & 6.59    & 6.07   & 5.71   & 6.80   \\ \hdashline 
Load Instructions        & 3.51  & 3.04  & 2.85  & 2.94  & 1.00    & 1.06   & 0.90   & 1.27   \\ 
Jump Instructions        & 5.21  & 4.46  & 4.33  & 5.01  & 7.51    & 7.65   & 7.04   & 9.14   \\
Total Instructions       & 69.46 & 48.05 & 49.71 & 55.99 & 78.13   & 50.57  & 52.30  & 70.38  \\ \bottomrule
\end{tabular}
  \caption{This table lists the average number of different labels (jump labels, data labels) and instructions (jump instructions, load instructions, total instructions) contained in each sample across various compilation optimization levels (O0 to O3) within the Exebench and Decompile-Eval datasets.}
  \label{tab:dataset}
\end{table}

%% file: sections/related.tex
\section{BACKGROUND \& RELATED WORK}
\label{sec:rel}

Decompilation is the process of reversing a binary file back into its source code form.
This process can be used to analyze the functionality of software when the source code is unavailable.
The typical decompilation tools, such as Hex-Rays IDA Pro~\citep{idapro} and Ghidra~\citep{ghidra}, typically rely on the analysis of the program's data flow or control flow~\citep{decompilation1}.
These decompilation tools analyze the instructions in the executable section (.text section) of the assembly code and then construct the program's Control Flow Graph (CFG). 
They identify patterns that correspond to standard programming structures (such as if-else, while loops, etc.~\citep{for_loop}) and perform type inference to resolve information in the read-only data section (.rodata section).

However, the construction of these tools heavily relies on rule systems created by experts, and the process of constructing these rules is highly challenging.
It is also difficult to cover the entire CFG, and errors are common.
Furthermore, these rules tend to fail when facing optimized binary code, yet optimization is a common practice in commercial compilers.
Additionally, the output of these decompilation tools is often a source-code-like representation of the assembly code, such as directly translating variables to registers, using goto statements, and other low-level operations.
This makes the output code difficult to read and understand, and it may not be sufficient to support recompilation.

Inspired by neural machine translation, researchers redefine decompilation as a translation task, which converts machine-level instructions into human-readable source code.
Initial attempts use Recurrent Neural Networks (RNNs)~\citep{decompilation2_rnn} for decompilation, supplemented by error correction techniques to improve results.
However, these efforts are limited in effectiveness.
Recent advancements in Natural Language Processing (NLP) enable large language models (LLMs) to be applied to code-related tasks~\citep{codellama, starcoder, deepseekcoder}. 
These models typically adopt the Transformer architectures~\citep{transformers}, use self-attention mechanisms, and are pre-trained on large-scale text datasets.
This approach allows LLMs to capture subtle contextual nuances and contributes to a general understanding of language.
Currently, when introducing LLMs into the binary decompilation domain, the methods are categorized based on whether relying on existing decompilation tools: end-to-end methods and refine-based methods. 
We will respectively introduce them below.

Specifically, refine-based methods work with the output of decompilation tools.
DeGPT\citep{hu2024degpt} designs an end-to-end framework to improve the readability of decompiler output.
DecGPT\citep{refine_decompile} uses LLMs combined with compiler information and runtime program information to enhance the compilability of decompiler output.
Recently, LLM4Decompile~\citep{llm4decompile} releases the first open-source large language model specifically for decompilation, which includes both refine-based models and end-to-end models.
Refine-based methods reuse human-encoded rules from existing decompilers, reducing the difficulty of decompilation, but they also introduce additional dependencies.

End-to-end methods decompile directly from assembly code.
BTC~\citep{btc} is one of the earliest methods to fine-tune LLMs for this purpose, extending the decompilation task to multiple languages.
Slade~\citep{slade} expand the model size and trained an LLM-based decompiler with 200 million parameters.
Nova~\citep{nova} proposes a hierarchical attention mechanism and a contrastive learning approach to improve the decompilation ability of the model.
LLM4Decompile\citep{llm4decompile}, in addition to releasing open-source decompilation models, provides a new benchmark for decompilation tasks.
SCC and FAE \citep{feng2024self} further improve the performance of end-to-end decompilation based on LLM4Decompile with self-constructed context and fine-grained alignment techniques.

We find that existing end-to-end decompilation methods perform poorly in handling binary files.
They not only lose the jump information in the executable section but also fail to retain the information from the data sections needed for decompilation.
This could be a key reason for the poor performance of end-to-end methods.
To address the challenges, we redesign the end-to-end decompilation process in \textit{ReF Decompile}, incorporating relabeling strategy to preserve control flow information and function call strategy to access variable information.

%% file: sections/conclusion.tex
\section{CONCLUSION}

Decompilation is the reverse process of converting compiled binary code back into a high-level programming language. 
In this paper, we revisit previous end-to-end decompilation approaches and identify a critical issue: they often lose crucial information required for reconstructing control flow structures and restoring variables when processing binary files.
This limitation makes it challenging for end-to-end methods to accurately recover program logic, resulting in inferior performance compared to refine-based methods.
To address this issue, we propose \textbf{ReF Decompile}, which involves a redesigned end-to-end decompilation workflow.
Specifically, to tackle the loss of control flow information, we introduce a relabeling strategy to reformat data by replacing jump target addresses with labels and placing the corresponding labels before the jump targets for clear identification.
To mitigate the loss of variable information, we train the model to infer variable types using function call strategy, allowing interaction with the binary file to retrieve variable values and complete the information required for variable reconstruction.
Experimental results on the Humaneval-Decompile Benchmark demonstrate that ReF Decompile, as an end-to-end approach, outperforms refine-based baselines with the same model size.
It achieves a SOTA performance of 61.43\% in deep learning decompilation.
In addition, we find that our method not only enhances the performance of the decompilation task but also improves the readability of the decompiled results compared to baselines.
We further analyze the effectiveness of the Relabeling and Function Call through ablation studies and dataset analysis.

%% file: colm_conference.bbl
\begin{thebibliography}{28}
\providecommand{\natexlab}[1]{#1}
\providecommand{\url}[1]{\texttt{#1}}
\expandafter\ifx\csname urlstyle\endcsname\relax
  \providecommand{\doi}[1]{doi: #1}\else
  \providecommand{\doi}{doi: \begingroup \urlstyle{rm}\Url}\fi

\bibitem[Armengol-Estap\'{e} et~al.(2022)Armengol-Estap\'{e}, Woodruff, Brauckmann, Magalh\~{a}es, and O'Boyle]{exebench}
Jordi Armengol-Estap\'{e}, Jackson Woodruff, Alexander Brauckmann, Jos\'{e} Wesley de~Souza Magalh\~{a}es, and Michael F.~P. O'Boyle.
\newblock Exebench: An ml-scale dataset of executable c functions.
\newblock In \emph{Proceedings of the 6th ACM SIGPLAN International Symposium on Machine Programming}, MAPS 2022, pp.\  50–59, New York, NY, USA, 2022. Association for Computing Machinery.
\newblock ISBN 9781450392730.
\newblock \doi{10.1145/3520312.3534867}.
\newblock URL \url{https://doi.org/10.1145/3520312.3534867}.

\bibitem[Armengol{-}Estap{\'{e}} et~al.(2023)Armengol{-}Estap{\'{e}}, Woodruff, Cummins, and O'Boyle]{slade}
Jordi Armengol{-}Estap{\'{e}}, Jackson Woodruff, Chris Cummins, and Michael F.~P. O'Boyle.
\newblock Slade: {A} portable small language model decompiler for optimized assembler.
\newblock \emph{CoRR}, abs/2305.12520, 2023.
\newblock \doi{10.48550/ARXIV.2305.12520}.
\newblock URL \url{https://doi.org/10.48550/arXiv.2305.12520}.

\bibitem[Brumley et~al.(2013)Brumley, Lee, Schwartz, and Woo]{decompilation1}
David Brumley, JongHyup Lee, Edward~J. Schwartz, and Maverick Woo.
\newblock Native x86 decompilation using semantics-preserving structural analysis and iterative control-flow structuring.
\newblock In Samuel~T. King (ed.), \emph{Proceedings of the 22th {USENIX} Security Symposium, Washington, DC, USA, August 14-16, 2013}, pp.\  353--368. {USENIX} Association, 2013.
\newblock URL \url{https://www.usenix.org/conference/usenixsecurity13/technical-sessions/presentation/schwartz}.

\bibitem[Chen et~al.(2021)Chen, Tworek, Jun, Yuan, de~Oliveira~Pinto, Kaplan, Edwards, Burda, Joseph, Brockman, Ray, Puri, Krueger, Petrov, Khlaaf, Sastry, Mishkin, Chan, Gray, Ryder, Pavlov, Power, Kaiser, Bavarian, Winter, Tillet, Such, Cummings, Plappert, Chantzis, Barnes, Herbert{-}Voss, Guss, Nichol, Paino, Tezak, Tang, Babuschkin, Balaji, Jain, Saunders, Hesse, Carr, Leike, Achiam, Misra, Morikawa, Radford, Knight, Brundage, Murati, Mayer, Welinder, McGrew, Amodei, McCandlish, Sutskever, and Zaremba]{humaneval}
Mark Chen, Jerry Tworek, Heewoo Jun, Qiming Yuan, Henrique~Pond{\'{e}} de~Oliveira~Pinto, Jared Kaplan, Harrison Edwards, Yuri Burda, Nicholas Joseph, Greg Brockman, Alex Ray, Raul Puri, Gretchen Krueger, Michael Petrov, Heidy Khlaaf, Girish Sastry, Pamela Mishkin, Brooke Chan, Scott Gray, Nick Ryder, Mikhail Pavlov, Alethea Power, Lukasz Kaiser, Mohammad Bavarian, Clemens Winter, Philippe Tillet, Felipe~Petroski Such, Dave Cummings, Matthias Plappert, Fotios Chantzis, Elizabeth Barnes, Ariel Herbert{-}Voss, William~Hebgen Guss, Alex Nichol, Alex Paino, Nikolas Tezak, Jie Tang, Igor Babuschkin, Suchir Balaji, Shantanu Jain, William Saunders, Christopher Hesse, Andrew~N. Carr, Jan Leike, Joshua Achiam, Vedant Misra, Evan Morikawa, Alec Radford, Matthew Knight, Miles Brundage, Mira Murati, Katie Mayer, Peter Welinder, Bob McGrew, Dario Amodei, Sam McCandlish, Ilya Sutskever, and Wojciech Zaremba.
\newblock Evaluating large language models trained on code.
\newblock \emph{CoRR}, abs/2107.03374, 2021.
\newblock URL \url{https://arxiv.org/abs/2107.03374}.

\bibitem[Dao(2024)]{flashattention2}
Tri Dao.
\newblock Flash{A}ttention-2: Faster attention with better parallelism and work partitioning.
\newblock In \emph{International Conference on Learning Representations (ICLR)}, 2024.

\bibitem[Feng et~al.(2024)Feng, Teng, Xu, Mu, Xu, Qin, Zhu, and Che]{feng2024self}
Yunlong Feng, Dechuan Teng, Yang Xu, Honglin Mu, Xiao Xu, Libo Qin, Qingfu Zhu, and Wanxiang Che.
\newblock Self-constructed context decompilation with fined-grained alignment enhancement.
\newblock In Yaser Al-Onaizan, Mohit Bansal, and Yun-Nung Chen (eds.), \emph{Findings of the Association for Computational Linguistics: EMNLP 2024}, pp.\  6603--6614, Miami, Florida, USA, November 2024. Association for Computational Linguistics.
\newblock URL \url{https://aclanthology.org/2024.findings-emnlp.385}.

\bibitem[Ghidra(2024)]{ghidra}
Ghidra.
\newblock Ghidra software reverse engineering framework, 2024.
\newblock URL \url{https://github.com/NationalSecurityAgency/ghidra}.

\bibitem[Guo et~al.(2024)Guo, Zhu, Yang, Xie, Dong, Zhang, Chen, Bi, Wu, Li, Luo, Xiong, and Liang]{deepseekcoder}
Daya Guo, Qihao Zhu, Dejian Yang, Zhenda Xie, Kai Dong, Wentao Zhang, Guanting Chen, Xiao Bi, Y.~Wu, Y.~K. Li, Fuli Luo, Yingfei Xiong, and Wenfeng Liang.
\newblock Deepseek-coder: When the large language model meets programming -- the rise of code intelligence, 2024.

\bibitem[Hex-Rays(2024)]{idapro}
Hex-Rays.
\newblock Ida pro: a cross-platform multi-processor disassembler and debugger, 2024.
\newblock URL \url{https://hex-rays.com/ida-pro/}.

\bibitem[Hosseini \& Dolan{-}Gavitt(2022)Hosseini and Dolan{-}Gavitt]{btc}
Iman Hosseini and Brendan Dolan{-}Gavitt.
\newblock Beyond the {C:} retargetable decompilation using neural machine translation.
\newblock \emph{CoRR}, abs/2212.08950, 2022.
\newblock \doi{10.48550/ARXIV.2212.08950}.
\newblock URL \url{https://doi.org/10.48550/arXiv.2212.08950}.

\bibitem[Hu et~al.(2022)Hu, Shen, Wallis, Allen-Zhu, Li, Wang, Wang, and Chen]{hu2022lora}
Edward~J Hu, Yelong Shen, Phillip Wallis, Zeyuan Allen-Zhu, Yuanzhi Li, Shean Wang, Lu~Wang, and Weizhu Chen.
\newblock Lo{RA}: Low-rank adaptation of large language models.
\newblock In \emph{International Conference on Learning Representations}, 2022.
\newblock URL \url{https://openreview.net/forum?id=nZeVKeeFYf9}.

\bibitem[Hu et~al.(2024)Hu, Liang, and Chen]{hu2024degpt}
Peiwei Hu, Ruigang Liang, and Kai Chen.
\newblock Degpt: Optimizing decompiler output with llm.
\newblock In \emph{Proceedings 2024 Network and Distributed System Security Symposium (2024). https://api. semanticscholar. org/CorpusID}, volume 267622140, 2024.

\bibitem[Hui et~al.(2024)Hui, Yang, Cui, Yang, Liu, Zhang, Liu, Zhang, Yu, Lu, Dang, Fan, Zhang, Yang, Men, Huang, Zheng, Miao, Quan, Feng, Ren, Ren, Zhou, and Lin]{qwen2.5coder}
Binyuan Hui, Jian Yang, Zeyu Cui, Jiaxi Yang, Dayiheng Liu, Lei Zhang, Tianyu Liu, Jiajun Zhang, Bowen Yu, Keming Lu, Kai Dang, Yang Fan, Yichang Zhang, An~Yang, Rui Men, Fei Huang, Bo~Zheng, Yibo Miao, Shanghaoran Quan, Yunlong Feng, Xingzhang Ren, Xuancheng Ren, Jingren Zhou, and Junyang Lin.
\newblock Qwen2.5-coder technical report, 2024.
\newblock URL \url{https://arxiv.org/abs/2409.12186}.

\bibitem[Jiang et~al.(2023)Jiang, Wang, Liu, Xu, Tan, and Zhang]{nova}
Nan Jiang, Chengxiao Wang, Kevin Liu, Xiangzhe Xu, Lin Tan, and Xiangyu Zhang.
\newblock Nova$^+$: Generative language models for binaries, 2023.

\bibitem[Katz et~al.(2018)Katz, Ruchti, and Schulte]{decompilation2_rnn}
Deborah~S. Katz, Jason Ruchti, and Eric~M. Schulte.
\newblock Using recurrent neural networks for decompilation.
\newblock In Rocco Oliveto, Massimiliano~Di Penta, and David~C. Shepherd (eds.), \emph{25th International Conference on Software Analysis, Evolution and Reengineering, {SANER} 2018, Campobasso, Italy, March 20-23, 2018}, pp.\  346--356. {IEEE} Computer Society, 2018.
\newblock \doi{10.1109/SANER.2018.8330222}.
\newblock URL \url{https://doi.org/10.1109/SANER.2018.8330222}.

\bibitem[Kirchner \& Rosenthaler(2017)Kirchner and Rosenthaler]{decompile_ir1}
Kevin Kirchner and Stefan Rosenthaler.
\newblock bin2llvm: Analysis of binary programs using {LLVM} intermediate representation.
\newblock In \emph{Proceedings of the 12th International Conference on Availability, Reliability and Security, Reggio Calabria, Italy, August 29 - September 01, 2017}, pp.\  45:1--45:7. {ACM}, 2017.
\newblock \doi{10.1145/3098954.3103152}.
\newblock URL \url{https://doi.org/10.1145/3098954.3103152}.

\bibitem[Lacomis et~al.(2019)Lacomis, Yin, Schwartz, Allamanis, Goues, Neubig, and Vasilescu]{variable_name}
Jeremy Lacomis, Pengcheng Yin, Edward~J. Schwartz, Miltiadis Allamanis, Claire~Le Goues, Graham Neubig, and Bogdan Vasilescu.
\newblock {DIRE:} {A} neural approach to decompiled identifier naming.
\newblock In \emph{34th {IEEE/ACM} International Conference on Automated Software Engineering, {ASE} 2019, San Diego, CA, USA, November 11-15, 2019}, pp.\  628--639. {IEEE}, 2019.
\newblock \doi{10.1109/ASE.2019.00064}.
\newblock URL \url{https://doi.org/10.1109/ASE.2019.00064}.

\bibitem[Lippincott(2020)]{starcoder}
Thomas Lippincott.
\newblock Starcoder: {A} general neural ensemble technique to support traditional scholarship, illustrated with a study of the post-atlantic slave trade.
\newblock In Laura Estill and Jennifer Guiliano (eds.), \emph{15th Annual International Conference of the Alliance of Digital Humanities Organizations, {DH} 2020, Ottawa, Canada, July 20-25, 2020, Conference Abstracts}, 2020.

\bibitem[Loshchilov \& Hutter(2019)Loshchilov and Hutter]{adamw}
Ilya Loshchilov and Frank Hutter.
\newblock Decoupled weight decay regularization.
\newblock In \emph{7th International Conference on Learning Representations, {ICLR} 2019, New Orleans, LA, USA, May 6-9, 2019}. OpenReview.net, 2019.
\newblock URL \url{https://openreview.net/forum?id=Bkg6RiCqY7}.

\bibitem[OpenAI(2023)]{gpt4}
OpenAI.
\newblock {GPT-4} {T}echnical {R}eport.
\newblock \emph{CoRR}, abs/2303.08774, 2023.
\newblock \doi{10.48550/arXiv.2303.08774}.
\newblock URL \url{https://doi.org/10.48550/arXiv.2303.08774}.

\bibitem[Rozi{\`{e}}re et~al.(2023)Rozi{\`{e}}re, Gehring, Gloeckle, Sootla, Gat, Tan, Adi, Liu, Remez, Rapin, Kozhevnikov, Evtimov, Bitton, Bhatt, Canton{-}Ferrer, Grattafiori, Xiong, D{\'{e}}fossez, Copet, Azhar, Touvron, Martin, Usunier, Scialom, and Synnaeve]{codellama}
Baptiste Rozi{\`{e}}re, Jonas Gehring, Fabian Gloeckle, Sten Sootla, Itai Gat, Xiaoqing~Ellen Tan, Yossi Adi, Jingyu Liu, Tal Remez, J{\'{e}}r{\'{e}}my Rapin, Artyom Kozhevnikov, Ivan Evtimov, Joanna Bitton, Manish Bhatt, Cristian Canton{-}Ferrer, Aaron Grattafiori, Wenhan Xiong, Alexandre D{\'{e}}fossez, Jade Copet, Faisal Azhar, Hugo Touvron, Louis Martin, Nicolas Usunier, Thomas Scialom, and Gabriel Synnaeve.
\newblock Code llama: Open foundation models for code.
\newblock \emph{CoRR}, abs/2308.12950, 2023.
\newblock \doi{10.48550/ARXIV.2308.12950}.
\newblock URL \url{https://doi.org/10.48550/arXiv.2308.12950}.

\bibitem[Tan et~al.(2024)Tan, Luo, Li, and Zhang]{llm4decompile}
Hanzhuo Tan, Qi~Luo, Jing Li, and Yuqun Zhang.
\newblock {LLM}4{D}ecompile: Decompiling binary code with large language models.
\newblock In Yaser Al-Onaizan, Mohit Bansal, and Yun-Nung Chen (eds.), \emph{Proceedings of the 2024 Conference on Empirical Methods in Natural Language Processing}, pp.\  3473--3487, Miami, Florida, USA, November 2024. Association for Computational Linguistics.
\newblock URL \url{https://aclanthology.org/2024.emnlp-main.203}.

\bibitem[Vaswani et~al.(2017)Vaswani, Shazeer, Parmar, Uszkoreit, Jones, Gomez, Kaiser, and Polosukhin]{transformers}
Ashish Vaswani, Noam Shazeer, Niki Parmar, Jakob Uszkoreit, Llion Jones, Aidan~N. Gomez, Lukasz Kaiser, and Illia Polosukhin.
\newblock Attention is all you need.
\newblock In Isabelle Guyon, Ulrike von Luxburg, Samy Bengio, Hanna~M. Wallach, Rob Fergus, S.~V.~N. Vishwanathan, and Roman Garnett (eds.), \emph{Advances in Neural Information Processing Systems 30: Annual Conference on Neural Information Processing Systems 2017, December 4-9, 2017, Long Beach, CA, {USA}}, pp.\  5998--6008, 2017.
\newblock URL \url{https://proceedings.neurips.cc/paper/2017/hash/3f5ee243547dee91fbd053c1c4a845aa-Abstract.html}.

\bibitem[Wang et~al.(2024)Wang, Qin, Ade~Jacobs, Holmes, Rajbhandari, Ruwase, Yang, Yang, and He]{wang2024zero}
Guanhua Wang, Heyang Qin, Sam Ade~Jacobs, Connor Holmes, Samyam Rajbhandari, Olatunji Ruwase, Feng Yang, Lei Yang, and Yuxiong He.
\newblock Zero++: Extremely efficient collective communication for giant model training.
\newblock In \emph{ICLR 2024}, May 2024.
\newblock URL \url{https://www.microsoft.com/en-us/research/publication/zero-extremely-efficient-collective-communication-for-giant-model-training/}.

\bibitem[Wei et~al.(2007)Wei, Mao, Zou, and Chen]{for_loop}
Tao Wei, Jian Mao, Wei Zou, and Yu~Chen.
\newblock A new algorithm for identifying loops in decompilation.
\newblock In Hanne~Riis Nielson and Gilberto Fil{\'{e}} (eds.), \emph{Static Analysis, 14th International Symposium, {SAS} 2007, Kongens Lyngby, Denmark, August 22-24, 2007, Proceedings}, volume 4634 of \emph{Lecture Notes in Computer Science}, pp.\  170--183. Springer, 2007.
\newblock \doi{10.1007/978-3-540-74061-2\_11}.
\newblock URL \url{https://doi.org/10.1007/978-3-540-74061-2\_11}.

\bibitem[Wolf et~al.(2019)Wolf, Debut, Sanh, Chaumond, Delangue, Moi, Cistac, Rault, Louf, Funtowicz, and Brew]{huggingface}
Thomas Wolf, Lysandre Debut, Victor Sanh, Julien Chaumond, Clement Delangue, Anthony Moi, Pierric Cistac, Tim Rault, R{\'{e}}mi Louf, Morgan Funtowicz, and Jamie Brew.
\newblock Huggingface's transformers: State-of-the-art natural language processing.
\newblock \emph{CoRR}, abs/1910.03771, 2019.
\newblock URL \url{http://arxiv.org/abs/1910.03771}.

\bibitem[Wong et~al.(2023)Wong, Wang, Li, Liu, Wang, Tang, Nie, and Wu]{refine_decompile}
Wai~Kin Wong, Huaijin Wang, Zongjie Li, Zhibo Liu, Shuai Wang, Qiyi Tang, Sen Nie, and Shi Wu.
\newblock Refining decompiled {C} code with large language models.
\newblock \emph{CoRR}, abs/2310.06530, 2023.
\newblock \doi{10.48550/ARXIV.2310.06530}.
\newblock URL \url{https://doi.org/10.48550/arXiv.2310.06530}.

\bibitem[Zheng et~al.(2024)Zheng, Zhang, Zhang, Ye, Luo, and Ma]{llamafactory}
Yaowei Zheng, Richong Zhang, Junhao Zhang, Yanhan Ye, Zheyan Luo, and Yongqiang Ma.
\newblock Llamafactory: Unified efficient fine-tuning of 100+ language models.
\newblock \emph{arXiv preprint arXiv:2403.13372}, 2024.
\newblock URL \url{http://arxiv.org/abs/2403.13372}.

\end{thebibliography}
